\documentclass[aps,prev,twocolumn,preprintnumbers,floatfix,nofootinbib]{revtex4-1}
\pdfoutput=1

\usepackage{algorithm}
\usepackage{soul}
\usepackage{mathrsfs}
\usepackage{amsmath, amsthm, amssymb}
\usepackage{color}
\usepackage{epstopdf}
\usepackage[pdftex]{graphicx}
\usepackage{amssymb}
\usepackage{verbatim}
\usepackage{cancel}
\usepackage{hyperref}
\usepackage{enumerate}

\usepackage[normalem]{ulem}

\newcommand{\kcite}[1]{\cite{#1}}
\newcommand{\keq}[1]{Eqn.~(\ref{#1})}
\newcommand{\kfig}[1]{Fig.~\ref{#1}}

\newcommand{\kap}[1]{App.~\ref{#1}}

\begin{document}

\preprint{KCL-PH-TH/2019-64}

\title{Black hole formation in relativistic Oscillaton collisions}

\author{James Y. Widdicombe$^{a}$}
\email{j.y.widdicombe@gmail.com}
\author{Thomas Helfer$^{a b}$}
\email{thomashelfer@live.de}
\author{Eugene A. Lim$^{a}$}
\email{eugene.a.lim@gmail.com}

\vspace{1cm}
\affiliation{
${}^a$ Theoretical Particle Physics and Cosmology Group, Physics Department, Kings College London, Strand, London WC2R 2LS, United Kingdom}
\affiliation{
${}^b$ Department of Physics \& Astronomy, Johns Hopkins University, Baltimore, MD 21218, USA }

\begin{abstract}
We investigate the physics of black hole formation from the head-on collisions of boosted equal mass Oscillatons (OS) in full numerical relativity, for both the cases where the OS have equal phases or are maximally off-phase (anti-phase). While unboosted OS collisions will form a BH as long as their initial compactness $\mathcal{C}\equiv GM/R$ is above a numerically determined critical value $\mathcal{C}>0.035$, we find that imparting a small initial boost counter-intuitively \emph{prevents} the formation of black holes even if $\mathcal{C}> 0.035$. If the boost is further increased, at very high boosts $\gamma>1/12\mathcal{C}$, BH formation occurs as predicted by the hoop conjecture. These two limits combine to form a  ``stability band'' where collisions result in either the OS ``passing through'' (equal phase)  or ``bouncing back'' (anti-phase), with a critical point occurring around  ${\cal C}\approx 0.07$. We argue that the existence of this stability band can be explained by the competition between the free fall and the interaction timescales of the collision.
\end{abstract}

\maketitle

\section{Introduction}

Astronomy has moved into a new golden era with the historic measurements of gravitational waves (GW) from the binary coalescence of black holes \cite{PhysRevLett.116.061102,PhysRevLett.116.241103,PhysRevLett.118.221101,2041-8205-851-2-L35,PhysRevLett.119.141101} and neutron stars \cite{2017PhRvL.119p1101A}. The detection of GW170817  further pushed our understanding with the first multimessenger detection of GWs and electromagnetic signals \cite{2041-8205-848-2-L13,2041-8205-848-2-L12}.  These signals have renewed the interest in the search for signals from exotic compact objects (ECO; see e.g. \cite{Alcubierre:2003sx,2016JCAP...10..001G,Cardoso:2016oxy,Hui:2016ltb,Palenzuela:2006wp,Brito:2015yfh,Hanna:2016uhs}) which are strongly gravitating objects which are made out of exotic matter. 

Self gravitating scalar field solitons are known to have highly compact cores \cite{LEE1992251,PhysRevLett.57.2485,PhysRevLett.66.1659} and provide a family of ECO candidates including Wheeler's ``geons'' \cite{PhysRev.97.511,RevModPhys.29.480}, boson stars \cite{PhysRev.172.1331}, and oscillatons \cite{1968PhRv..172.1331K,1969PhRv..187.1767R,Liddle:1993ha,1994PhRvL..72.2516S,1991PhRvL..66.1659S}. These are closely related to a family of objects known as axion stars \cite{Berezhiani:1989fu,Berezhiani:1989fp,Sakharov1994id,Berezhiani:1992rk,pecceiquinn1977,weinberg1978,wilczek1978,2014JHEP...06..037D,2010ARNPS..60..405J,2006JHEP...06..051S,axiverse,2006JHEP...05..078C,Marsh:2015xka,Cicoli:2012sz,2017JCAP...03..055H,Widdicombe:2018oeo}.

Recent work with scalar compact objects head on mergers \cite{Cardoso:2016oxy,Palenzuela:2006wp,Helfer:2018vtq,Choptuik:2009ww,Helfer:2018vtq} as well as mixed mergers \cite{Clough:2018exo,Dietrich:2018jov,Dietrich:2018bvi}, indicates distinctions in the gravitational wave signal with respect to black holes. If these distinctions also exist in binary coalescence (see \cite{Sennett:2017etc,Palenzuela:2017kcg,Bezares:2017mzk} for boson star inspirals), a single GW event could be a smoking gun for the existence of ECOs.

In this paper, we study the relativistic  head-on collisions of a class of real relativistic scalar fields solitons called oscillatons (OS) \cite{PhysRevLett.66.1659} using full (3+1) dimensional numerical relativity simulations with \textsc{GRChombo} \cite{Clough:2015sqa}. OS are stable on cosmological time scales \cite{Page:2003rd} and could be realised as an axion star where the leading order $\phi^4$ interaction is negligible due to having a high axion decay constant, $f_a$. Formation of such objects have been studied in both non-relativistic \cite{2017MNRAS.465..941D,Amin:2019ums} and relativistic cases \cite{Widdicombe:2018oeo}. 

One of the key features of an OS is that its scalar field configuration is not static. Instead it oscillates with the characteristic frequency $\omega\sim m$ where $m$ is the effective mass of the field which is inversely related to the axion decay constant $m\propto 1/f_a$. Thus the interactions of any pair of OS will depend not only on their respective masses and the geometry of the interactions, but also on their \emph{relative oscillation phase} $\Delta \theta$.

In the case of relativistic OS where gravity is strong, the OS can exhibit very high compactness on the order of tens of percent of the Schwarzschild radius.  In this regime, gravity back-reacts strongly on the configuration of the scalar field and sufficiently compact OS can interact to form black holes. In  \cite{Helfer:2018vtq}, we showed that the head-on collisions of unboosted OS in this regime can produce gravitational wave signals which are distinct and, at high compactness, more energetic than equivalent equal mass black hole mergers.

\begin{figure*}[ht!]
\centering
\includegraphics[width=2.08\columnwidth]{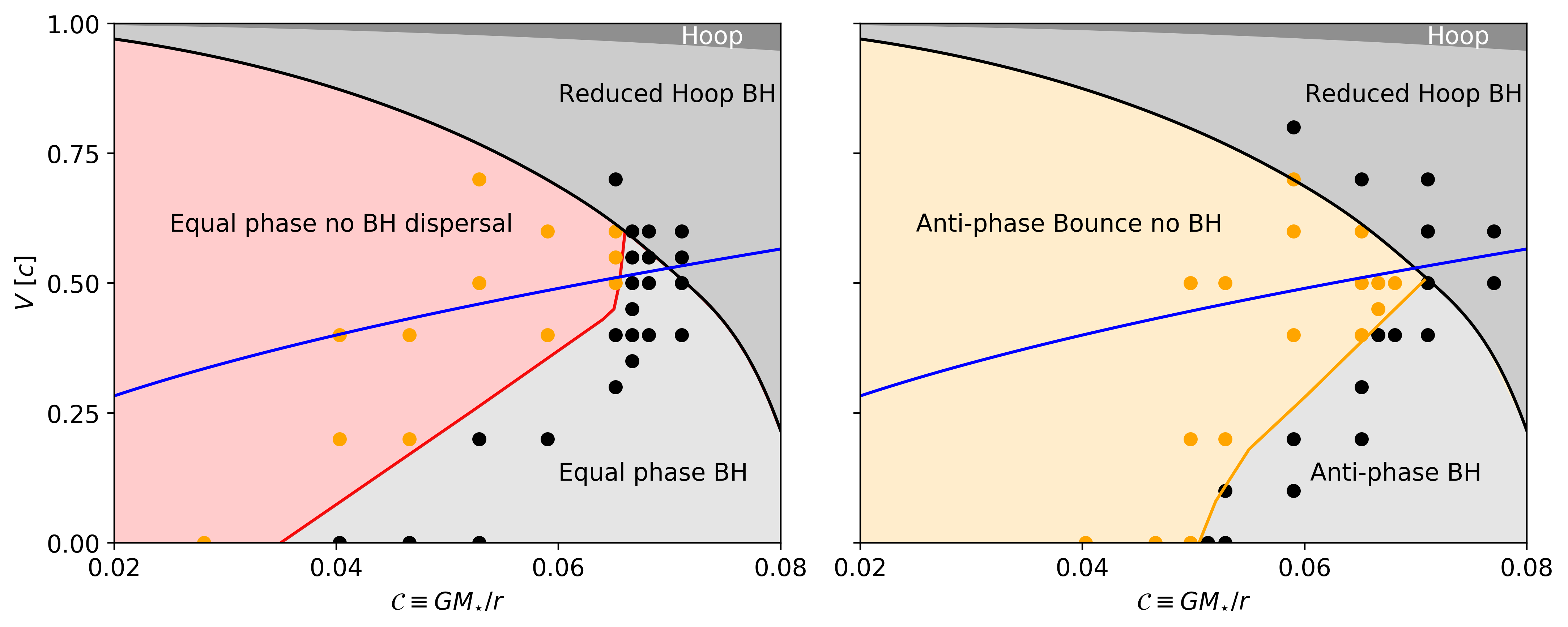}
  \caption{Final states of equal mass head-on OS-OS mergers as a function of compactness $\mathcal{C}$ and boost velocity $v$, for equal phase (left) and anti-phase cases (right). Dots indicate numerical simulations which end either in black hole formation (black) or dispersal/bounce (orange).  Shown are approximate regions indicating the final states of the collisions for the given initial conditions. The black line is the reduced hoop conjecture line \keq{eqn:hoop_2}, while the red (equal phase) and orange (anti-phase) lines are numerically determined estimates where above them black holes do not form.   In both cases, there exists a ``stability band'' between the black lines and the red/orange lines, in which the OS either disperse (equal phase) or bounce (anti-phase) post-collision. Comparing the free fall time and interaction times of the collision yields the blue line ($v \approx \mathcal{C}^{1/2}$), which converges with the reduced hoop conjecture line of $v\approx \sqrt{1-144\mathcal{C}^2}$  at $\mathcal{C}\approx 0.07$. 
}

\label{fig:money}
\end{figure*}

In this paper, we extend our work into two different directions. First, we consider the collisions of OS with different phases, in particular collisions in which their relative phase is maximal $\Delta \phi = \pi$, dubbed ``anti-phase'' OS collisions, confirming the perturbative gravity results of \cite{2016PDU....12...50P,PhysRevD.94.043513, Amin:2019ums}. We will show that anti-phase OS collisions experience a mutual repulsive force, confirming previous results obtained in perturbative gravity.  Secondly, we consider the collisions of \emph{boosted} OS, with  relativistic initial center of mass frame velocities, for both equal phase and anti-phase pairs of OS. While at high initial velocities, black holes formed as expected from the hoop conjecture argument \cite{Choptuik:2009ww,East:2012mb,Rezzolla:2012nr}, surprisingly and counter-intuitively, we show that at low velocities, collisions  are \emph{less likely to form black holes} when compared to the equivalent configuration with zero initial velocity. This effect is seen in both equal and anti-phase cases, indicating the possible existence of a ``critical point'' (see \kfig{fig:money}).

\section{Oscillatons and Initial set-up} \label{sect:oscillatons}

\begin{figure}
\centering
\includegraphics[width=1\columnwidth]{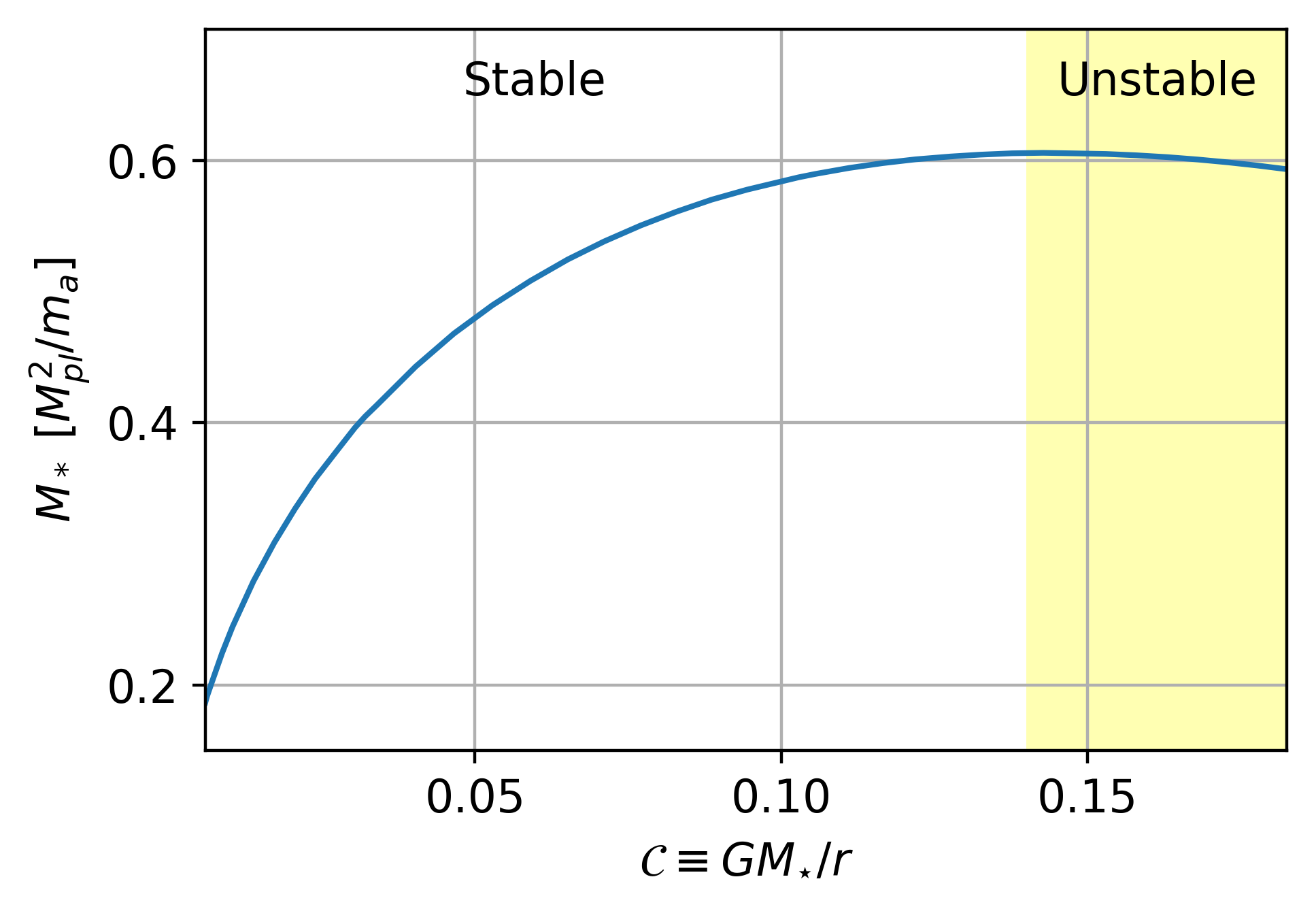}
\caption{Spherically symmetric unperturbed OS solutions are spanned by a single parameter, here chosen to be the compactness $\mathcal{C}=GM_*/R$, as found in \cite{Alcubierre:2003sx}. OS with $\mathcal{C}>0.14$ are unstable to perturbations, with perturbations either dissipating leading to a final state of $\mathcal{C}<0.14$ or collapsing into a black hole. }
\label{fig:compactness}
\end{figure}

We use units $\hbar=c=1$ and $M_{pl}=1/\sqrt{8\pi G_N}$ which is the reduced Planck mass.
Consider the action of a massive scalar field minimally coupled to gravity 
\begin{equation}
S = \int d^4 x \sqrt{-g} \left [  \frac{R}{16 \pi G} - \frac{1}{2} \partial_\mu \phi \partial^\mu \phi - \frac{1}{2} m^2 \phi^2 \right ] \,
\end{equation}
where $g$ is the determinant of the metric, $R$ is the Ricci scalar and $m$ is the mass of the real scalar field $\phi$. Such a potential suppports self-gravitating quasi-stable equilibrium OS \cite{PhysRevLett.66.1659}, and it has been shown  in \cite{Alcubierre:2003sx} that unexcited spherical symmetric solutions span a one-parameter family most conveniently represented by its compactness, $\mathcal{C}$, defined as
\begin{equation}
\mathcal{C} \equiv \frac{G M_*}{R} \,
\label{eqn:compactness}
\end{equation}
where $M_*$ is the total mass and $R$ is the radius. Note that for a given $\mathcal{C}$ the radius $R(M_*)$ of unexcited OS is completely determined by its mass $M_*$.  It has also been shown in \cite{Alcubierre:2003sx} that low compactness OS with $\mathcal{C} < 0.14$ are stable and typically migrate to other stable OS with $\mathcal{C}<0.14$ when strongly radially perturbed. On the other hand, high compactness OS with $\mathcal{C}>0.14$  are unstable, and under radial perturbations may either migrate to a stable lower mass OS with $\mathcal{C}<0.14$ via  scalar radiation  or collapse into a black hole (\kfig{fig:compactness}).

\begin{figure}
\includegraphics[width=1\columnwidth]{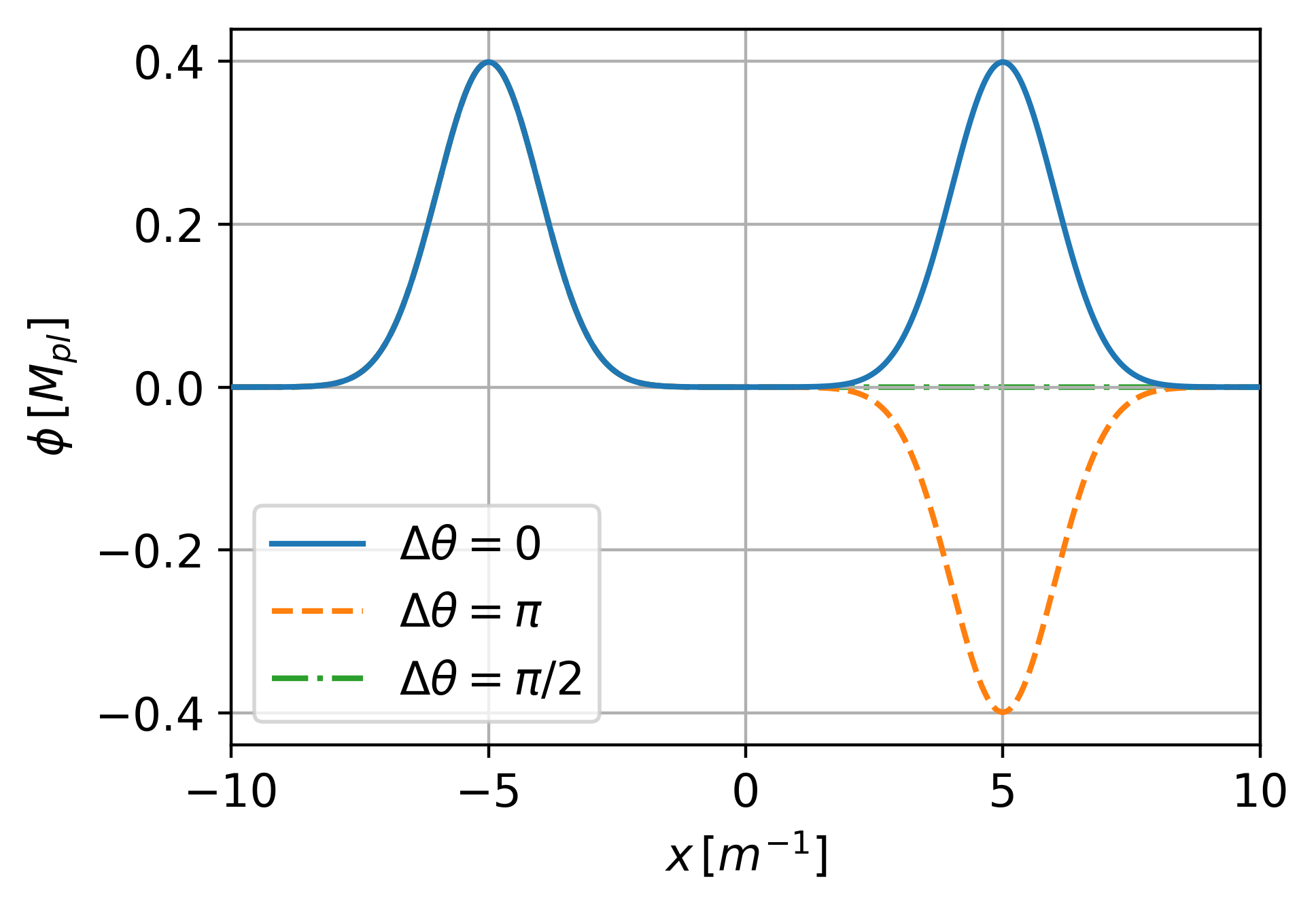}
\caption{One dimensional plot of the $\phi$ profile along the axis of collision of two OS for two different phases shown at fixed $t$ when the amplitude of $\phi$ for the left OS is maximised, with $x=0$ being the point of collision. The symmetry and anti-symmetry of the equal phase pair of OS ($\Delta \theta=0$) and an anti-phase pair of OS ($\Delta \theta=\pi$) respectively are constants of motion. }
\label{fig:relative_phase}
\end{figure}

A key property of OS is that it oscillates along a characteristic frequency $\omega \sim m$, and thus interactions of OS depend on their relative phase difference $\Delta \theta$. In particular, the field configuration $\phi(x,t)$ of a head-on collision of equal phase $\Delta \theta=0$ (anti-phase $\Delta \theta=\pi$) OS is \emph{symmetric} (\emph{anti-symmetric}) at the plane of collision parallel to the axis of motion. In between these two limits $0<\Delta \theta<\pi$, the collisions are said to be ``off-phase''. \kfig{fig:relative_phase} illustrates this further.

The special case for initially static, equal phase $\Delta \theta=0$ head-on collisions of OS was investigated in \cite{Helfer:2018vtq}. There, we showed that the end state of any such collision depends on the compactness $\mathcal{C}$. For $\mathcal{C}<0.035$ \emph{subcritical} collisions, the collision results in an excited more massive oscillaton, while for $0.035<\mathcal{C}< \mathcal{C}_*$ \emph{critical} collisions, the collision results in the formation of a black hole. For $\mathcal{C}>\mathcal{C}_*$  \emph{degenerate} collisions, since the OS are in the unstable branch (\kfig{fig:compactness}), mutual perturbations cause the OS to collapse into individual black holes before merging as a standard head on black hole collision.

In this paper, we will study both equal phase and anti-phase boosted head-on OS collisions.

\section{Boosted OS Collisions} \label{sect:boostedOS}

According to the hoop conjecture \cite{1972mwm..book..231T}, a quantity of matter/energy $E$ compressed into a spherical region such that a hoop of proper circumference $2\pi R$ completely encloses the matter in all directions, will form a black hole if the corresponding Schwarzschild radius, $R_s = 2 G E$ is greater then $R$.  The collisions of two solitons with individual rest mass $M_*$ boosted to $\gamma=(1-v^2)^{-1/2}$ will result in a system with an effective mass of $E=2\gamma M_*$ in the center of mass frame.  Applying the conjecture, if $R_s > R_0$ where $R_0$ is the rest frame radius of the soliton, then a black hole will form. Using \keq{eqn:compactness}, we obtain the following condition for black hole formation
\begin{equation}
\label{eqn:hoop}
\gamma \geq \frac{1}{4\mathcal{C}}~.
\end{equation}

Such relativistic collisions of scalar solitons have been studied numerically before in the context of ``boson stars\footnote{Boson stars are configurations of a complex scalar field with a $U(1)$ potential.  In contrast with the real scalar field OS which are stabilized by field oscillations,  boson stars are stabilized by their charges. For a review please see \cite{Liebling:2012fv}. }'' of $\mathcal{C}=0.025$ \cite{Choptuik:2009ww} and fluid packets  of $\mathcal{C}=0.0125$ \cite{East:2012mb}. In both cases, it was found that black hole formation occurs at the ``reduced'' hoop conjecture condition
\begin{equation}
\label{eqn:hoop_2}
\gamma \geq  \gamma_h \equiv \frac{1}{12\mathcal{C}}~,
\end{equation}
which is roughly about $1/3$ of what is predicted by the hoop conjecture. As we will soon see, we find this to be consistent with our simulations of relativistic OS collisions.

We simulated the collisions of two equal mass and hence equal $\mathcal{C}$ OS in numerical general relativity, using \textsc{GRChombo} \cite{Clough:2015sqa} for both equal phase and anti-phase cases. Their initial separation are set at $d=60m^{-1}$. We vary the initial velocities of the OS from $v=0$ to $v=0.8$ relative to the rest frame, with corresponding Lorentz factors $\gamma=1$ to $\gamma=1.4$ (see \kap{subsect:constructing_initial_data} for the details of the construction of initial data). In all cases except for $v=0$, the initial velocities are sufficiently high such that the OS are not initially bounded. 

We track the OS positions following \cite{Widdicombe:2018oeo} by locating the value and location of maximum density $\rho_{\mathrm{max}}$, which we identify as its center. While the OS started out initially spherical, during the collision process the OS becomes an ellipsoid due to the gravitational attraction along the axis of collision. The major and minor axes of the ellipsoid are then identified by the distance from the center to the point where the density is $5\%$ of $\rho_{\mathrm{center}}$.  Black hole formation is identified with a horizon finder. The results of our simulations is presented in Fig. \ref{fig:money}.

\subsection{Equal phase $\Delta \theta=0$ Collisions}

\begin{figure*}[ht!]
\centering
\includegraphics[width=2.0\columnwidth]{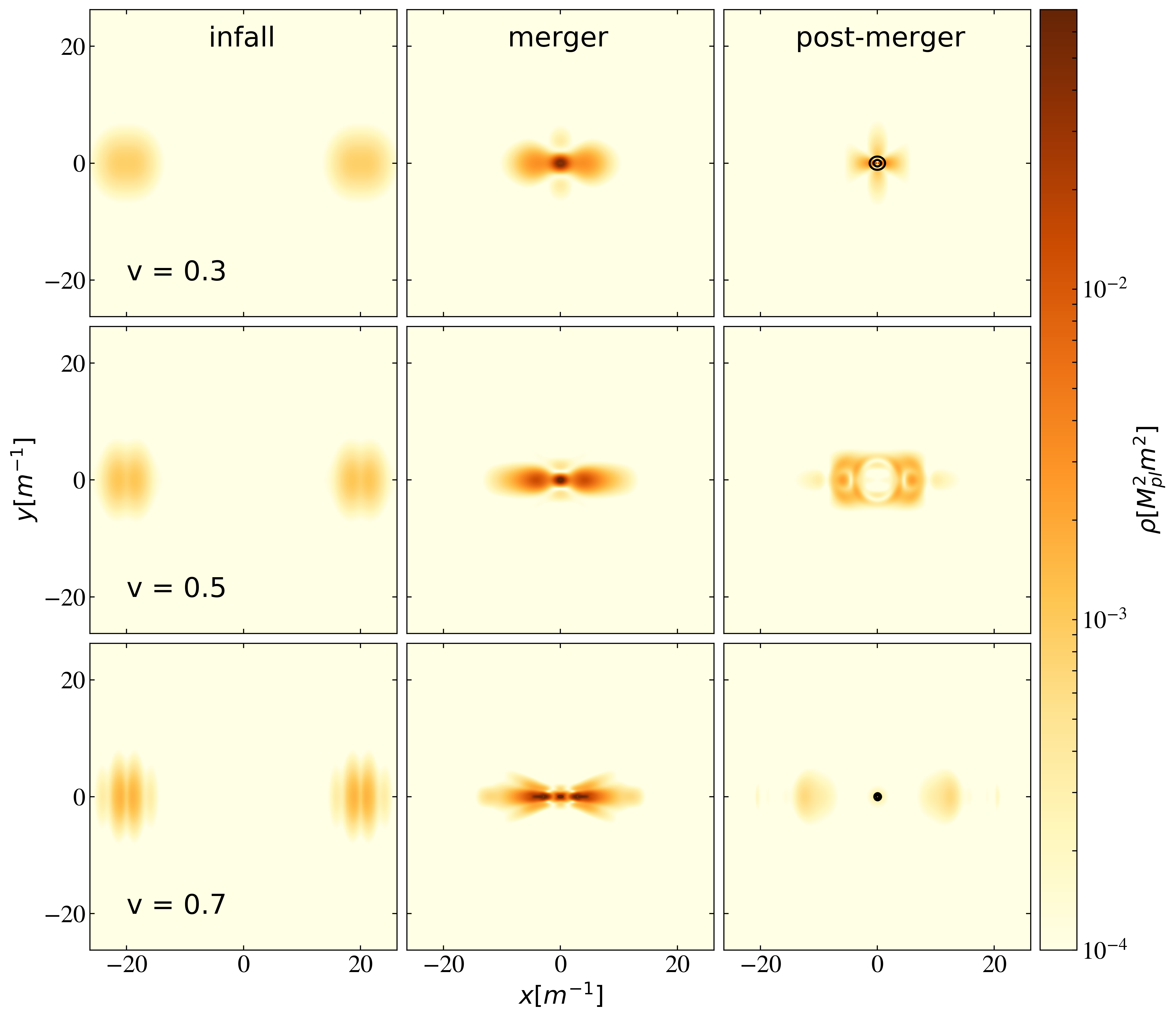}
\caption{{\bf In-phase $\Delta \theta = 0$ collisions :} Three different slices of energy density $\rho$ with $\mathcal{C} = 0.065$ with \href{https://youtu.be/mOPzPxIaDVg}{$v = 0.3~$}, \href{https://youtu.be/ZyYhJlYN3d8}{$~0.5~$},\href{https://youtu.be/66uwXSIY8tI}{$~ 0.7$} from top to bottom. The slices for the (i) infall, (ii) merger and (iii) post-merger. Black holes form  in the \href{https://youtu.be/mOPzPxIaDVg}{$v=0.3$} (top) and \href{https://youtu.be/66uwXSIY8tI}{$v=0.7$} (bottom) cases, with black lines indicating curvature contours at $\chi=0.2$ and $\chi=0.4$. In the \href{https://youtu.be/ZyYhJlYN3d8}{$v=0.5$} (middle) case, the  OS ``pass through'' each other and then dissipate. \href{https://www.youtube.com/playlist?list=PLSkfizpQDrcZJRY_vYHmp82OIfLwscNx8}{\color{blue}Link to movies} \cite{movieO3in,movieO5in,movieO7in}.} \label{fig:inphase} 
\end{figure*}

For equal phase $\Delta \theta =0$ case, at $v=0$ we recover the result of \cite{Helfer:2018vtq} whereby black hole formation occured when $\mathcal{C}\geq 0.035$.  At sufficiently high $v$, black holes form due to the additional energy imparted by the boost, as we expected. We found that they roughly obey the ``reduced''  hoop conjecture  argument \keq{eqn:hoop_2} (as opposed to \keq{eqn:hoop}), providing another data point to add to those of \cite{Choptuik:2009ww,East:2012mb, Rezzolla:2012nr}.

However, at low $v$, intriguingly, black hole formation occurs only at \emph{higher} compactness. For example, for $\mathcal{C}=0.04$, black holes will form at $v=0$ but will \emph{not} form at $v>0.2$ (until it meets the hoop conjecture line). In other words, \emph{initial non-zero velocities hinder the formation of black holes.} The velocity required to prevent black hole formation increases with increasing $\mathcal{C}$,  with the curve of transition sloping upwards until it meets the line defined by the  ``reduced'' hoop conjecture argument \keq{eqn:hoop_2}, at the ``critical'' point $\mathcal{C}\approx 0.068$ and $v\approx 0.55$.  Beyond this point $\mathcal{C}>0.068$, black holes form regardless of velocities.
In Fig. \ref{fig:inphase}, we show the black hole formation process of $\mathcal{C}=0.065$ OS collisions for the $v=0.7,~0.5,~0.3$ cases. 

The existence of this ``stability band'' for non-black hole end states can be explained by the fact that higher collisional velocities imply a shorter collision timescale. Since the boosted OS are not energetic enough to form black holes from the hoop conjecture alone, they must interact during the collision to form a sufficiently deep gravitational potential well to generate infall for a collapse into a black hole -- this defines an interaction/collapse timescale. However, in a sufficiently relativistic collision, the collision timescale may be shorter than the interaction/collapse timescale, resulting in the two OS ``passing through'' (or bouncing off) albeit with large perturbations to their initial configuration and at a slower velocity due to the inelastic nature of the collisions.

This collision timescale \emph{vs} interaction timescale behaviour has been seen in non-linear dynamics without gravity in the studies of relativistic collisions of non-linear solitons \cite{Giblin:2010bd,Amin:2013eqa,Amin:2013dqa}, where the relative coherence of the solitons post-collisions can be explained by the fact that the collision timescale is much shorter than the interaction timescale. We will discuss this in greater detail in the following  section \ref{sec:discussion}.

We find that the \emph{initial} formation of black holes is more efficient for the $v=0.3$ case when compared to the $v=0.7$ case -- the black hole mass grow more rapidly for the $v=0.3$ case during the collision. This could be due to the fact that the collision is ``messier'' when collisions are more energetic, and hence it takes longer for the excited debris to fall back into the nascent black hole. Unfortunately, our initial conditions are not sufficiently precise to enable long term tracking of the apparent horizon, leading to instabilities first seen in \cite{Okawa:2014nda}.

\subsection{Anti-phase $\Delta \theta=\pi$ Collisions}

\begin{figure*}[ht!]
\centering
\includegraphics[width=2.0\columnwidth]{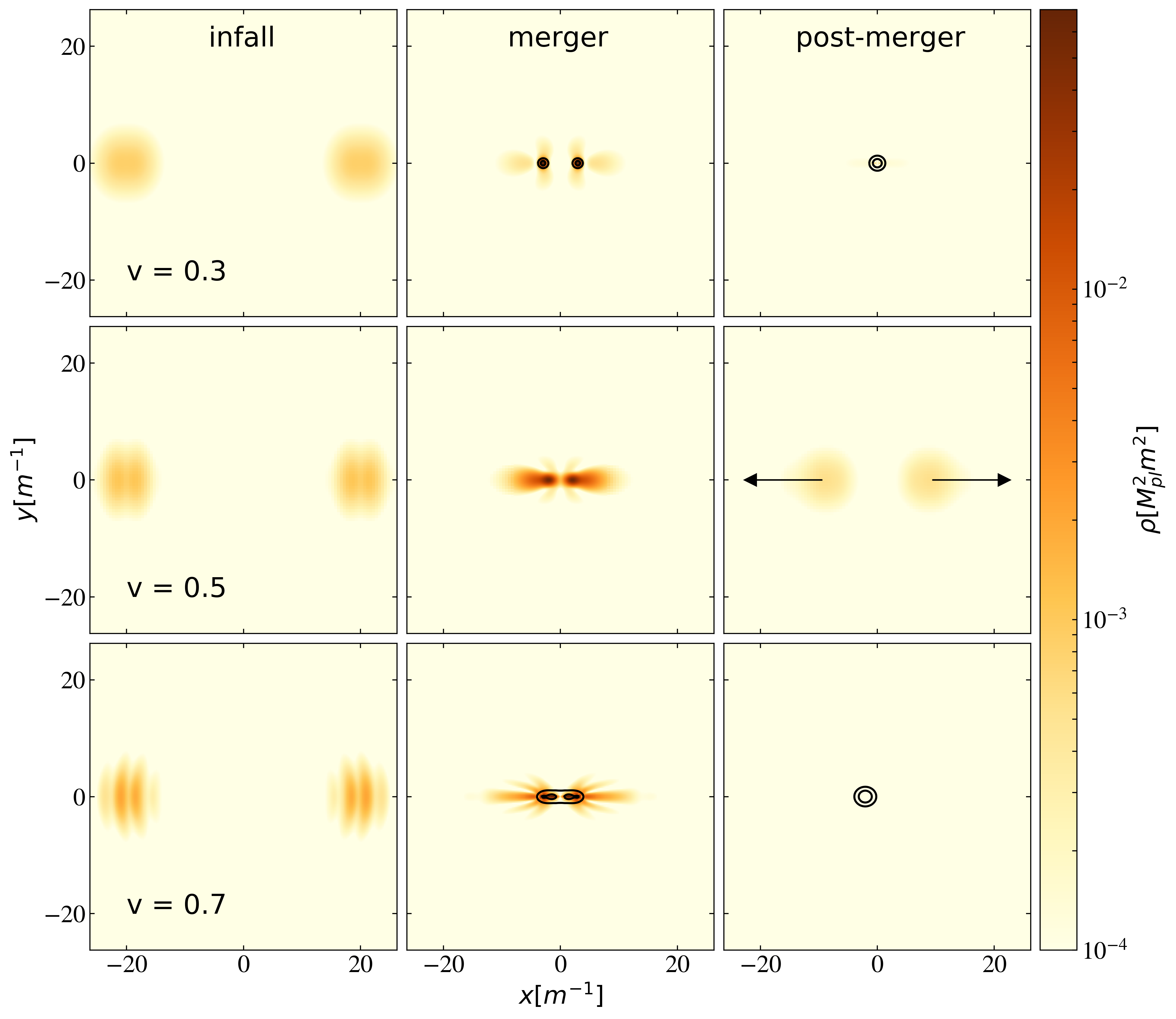}
\caption{{\bf Anti-phase $\Delta \theta = \pi$ collisions :} Three different slices of energy density $\rho$ with $\mathcal{C} = 0.065$ with \href{https://youtu.be/NyaB3zjtaQ4}{$v = 0.3~$},\href{https://youtu.be/uACT89NESHw}{$~0.5~$},\href{https://youtu.be/bdYYbXSgUcY}{$~ 0.7$} from top to bottom. The slices for the (i) infall, (ii) merger and (iii) post-merger. Black holes form  in the \href{https://youtu.be/NyaB3zjtaQ4}{$v=0.3$} (top) and \href{https://youtu.be/bdYYbXSgUcY}{$v=0.7$} (bottom) cases, with black lines indicating curvature contours at $\chi=0.2$ and $\chi=0.4$. In the \href{https://youtu.be/uACT89NESHw}{$v=0.5$} (middle) case, the  OS ``bounces back'' post-collision (with black arrows indicating the direction of travel). Notice that in both cases where black holes form, the OS collapse into black holes before merging. \href{https://www.youtube.com/watch?v=mOPzPxIaDVg&list=PLSkfizpQDrcZJRY_vYHmp82OIfLwscNx8}{\color{blue}Link to movies} \cite{movieO3off,movieO5off,movieO7off}.} \label{fig:offphase} 
\end{figure*}

At high $v$, black hole formation again occurs beyond the reduced hoop conjecture line \keq{eqn:hoop_2} -- reinforcing the point that in this regime ``matter does not matter'' and it is the gravitational dynamics that dominate \cite{Choptuik:2009ww}. Similar to the equal phase case above, at low $v$ black hole formation is impeded, although the transition line does not coincide, and is shifted slightly to the right (towards higher compactness). This line meets the reduced hoop conjecture line at the ``critical point''  $\mathcal{C}=0.071$ and $v=0.5$, indicating that there is an additional ``repulsion'' between the two OS when compared to the equal phase case. This repulsion is particularly notable in the $v=0$ case, where the transition from no black hole formation to black hole formation occurs at $\mathcal{C}\approx 0.05$ (compared to $\mathcal{C}\approx 0.035$ for equal phase collisions).

This repulsion can be explained as follows.  Crucially, for anti-phase collisions, the anti-symmetry of the $\phi$ configuration is a constant of motion, and hence at the point of collision $\phi(x_*,t)=0$ at all times where $x_*$ is the plane of anti-symmetry. This is in contrast with the equal phase pair where $\phi(x_*,t)$ is free to evolve as the two OS approach each other -- the symmetry of this case imposes the condition $\partial_x \phi(x_*,t)=0$ instead.  In particular, in \cite{2016PDU....12...50P,PhysRevD.94.043513, Amin:2019ums}, it was shown that in the weak gravity and non-relativistic limit, OS will ``bounce back'' instead of merging for $\Delta \phi<7\pi/8$ \cite{PhysRevD.94.043513}. In this limit, \cite{PhysRevD.94.043513} argues that since the oscillaton equation of motion is linear, in equal phase (anti-phase) collisions, the OS tend to constructively (destructively) interfere, at least at the collision plane $x_*$.

In strong gravity, gravitational back-reaction is non-linear, muddling this picture somewhat. Nevertheless, the anti-symmetry of the field configuration is still conserved, so $\phi(x_*,t)$ and its time derivative $\dot{\phi}(x_*,t)$ both remain at zero for all $t$. This means that the time averaged (over a period of oscillation) kinetic energy density of the field configuration $\langle E_K\rangle\sim (1/2)\dot{\phi}^2$ must vanish as $x\rightarrow x_*$.  As the OS approach each other, energy conservation forces the time averaged gradient energy $\langle E_G\rangle \sim (1/2)(\nabla \phi)^2$ to absorb this energy, resulting in a rapid increase in the gradient energy and thus a spiking of the scalar field spatial configuration\footnote{While it is natural to desribe this repulsion as a force, its behaviour is not described by a $1/r$ potential nor is it conservative. The anti-symmetric origin of the repulsion is reminiscent of the degenerate pressure of the anti-symmetric wavefunctions of fermions.} .  Note that the metric  and stress tensor remain symmetric in the diagonal components and anti-symmetric in the off-diagonal components throughout for both equal phase and anti-phase cases, which means that gravitational energy can still dominate near $x_*$.

\begin{figure}
\includegraphics[width=1\columnwidth]{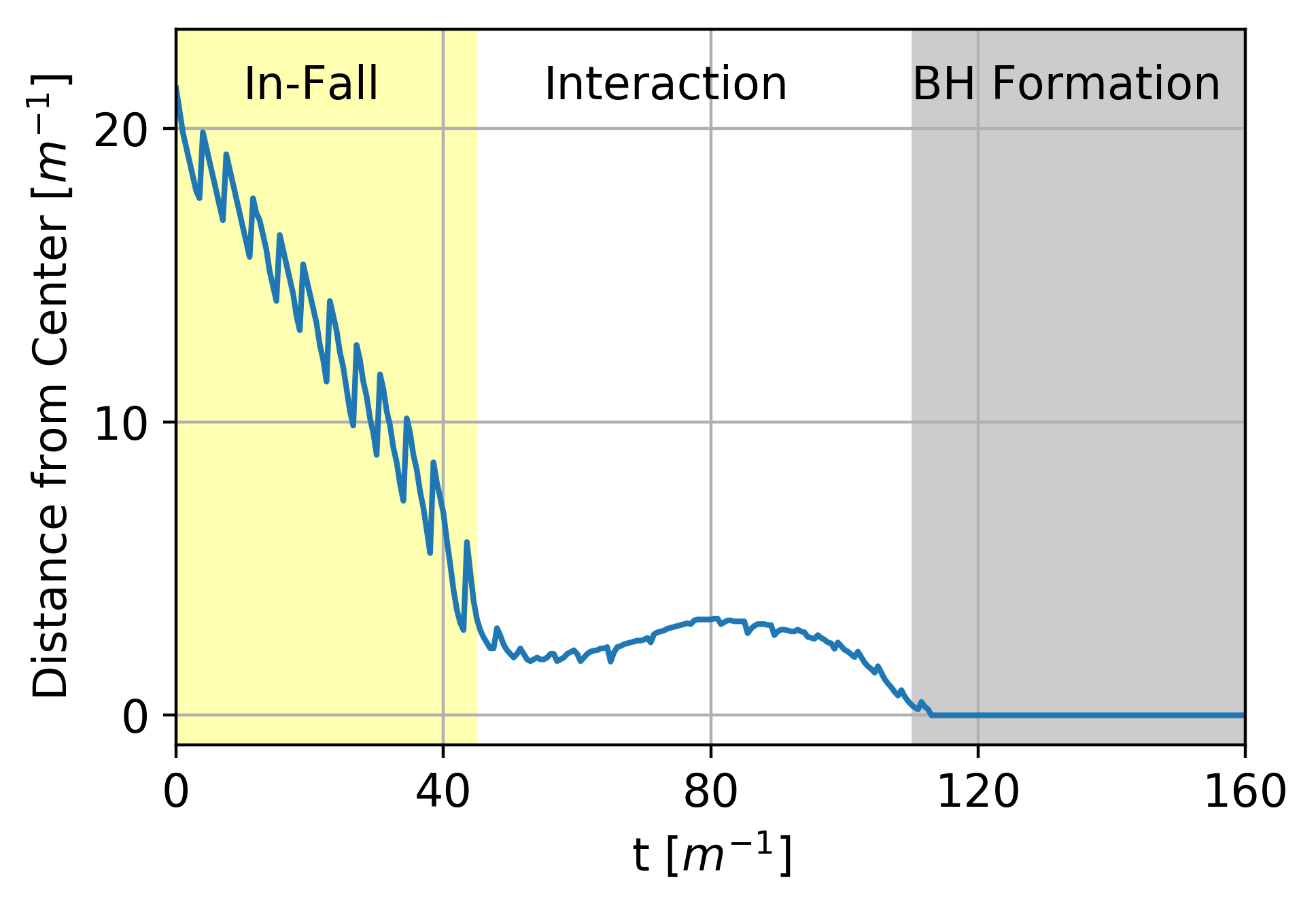}
\caption{The central location of a OS/BH vs time for an anti-phase OS collision with $\mathcal{C}=0.068$ and $v=0.4$. The repulsiveness of the anti-phase OS rapidly slows the initial velocity down to a full stop, before rebounding slightly at $t\sim 80 m^{-1}$ and then collapsing into a BH. The location of the center of the OS is taken to be the point of maximum density.}
\label{fig:rho_pos}
\end{figure}

\begin{figure}
\centering
\includegraphics[width=1\columnwidth]{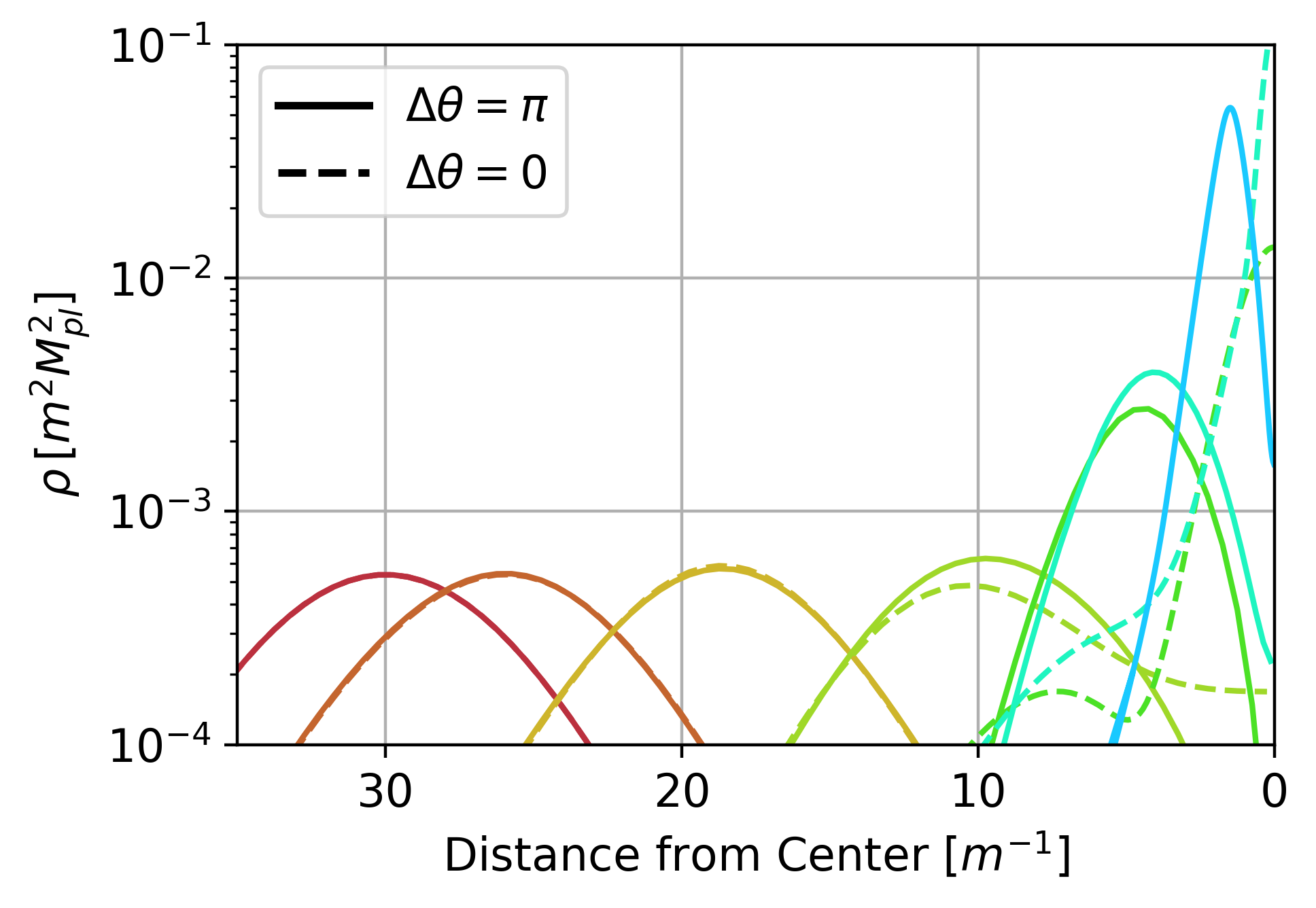}
\caption{
The time evolution of the profile of the energy density $\rho$ measured along the axis of collision for both equal phase (dotted line)  and anti-phase (continuous line)  collisions of OS with $\mathcal{C}= 0.053$. The time evolution is indicated by colour, chronologically increasing from deep red to blue. Note that anti-phase collisions experience a repulsion due to the anti-symmetry of the field configuration, and the centers (i.e. maximum density point)  of the OS remain distinct. As a result, the OS experience a compression which may lead to individual formation of black holes before final merger, or the OS ``bouncing back''.}
\label{fig:compression}
\end{figure}

To check this dependence, we ran a series of collisions with $\mathcal{C}= 0.028$ with zero boost for both OS, and an initial separation of $d=40m^{-1}$. For this compactness, we have previously shown in \cite{Helfer:2018vtq} that their mergers will lead to a highly excited OS in the limit of $\Delta \phi=0$, and hence we do not expect any black hole formation.

Since these are initially bound states, we expect that due to loss to scalar and gravitational wave radiation, the final state of such collisions will be a merged oscillaton. The key question is whether this merger occurs in the first collision as in the equal phase case, or will the off-phase repulsion generate pre-merger ``bounces''. We scan through $\Delta \phi = [0,\pi/8, \pi/4, 3\pi/8, \pi/2,5\pi/8, 3\pi/4, 7\pi/8, 15\pi/16, \pi]$, and found that only for the cases of $\Delta \phi \geq 7\pi/8$, the OS bounces once before merger -- in agreement with \cite{PhysRevD.94.043513} that this repulsion is only dominant when the phase difference is near maximal. The close agreement with the weak gravity results suggests that this repulsion effect is dominated by scalar dynamics.

\kfig{fig:compression} illustrates the comparison of the energy densities of equal phase and anti-phase collisions. At large distances, the two cases evolve similarly as they do not yet interact strongly. Their evolution begin to deviate around $d\sim 15m^{-1}$, as the OS begin to overlap and interact with each other. In the equal phase case, the OS merge and form a large central density spike at $d=0$. On the other hand, in the anti-phase case, the OS repulse each other -- note that the energy density drop at $d=0$ -- ``compressing'' to a smaller size but higher energy densities before bouncing back.

This repulsion and subsequent compression leads to a dramatically different black hole formation process when compared to the equal phase case. Instead of BH forming from the collapse of scalar matter after merger, the repulsion stops the motion of the OS and  prevents the direct merger of the OS from occurring. The accompanying  compression of both OS leads to a subsequent \emph{individual} collapse of the OS into separate black holes. These distinct black holes, shorn of the repulsive scalar field, then gravitate towards each other and finally form a final black hole.  This general mechanism is seen in both the high velocity (i.e. above the reduced hoop conjecture line) and low velocity BH formation processes (see Figs. \ref{fig:rho_pos} and \ref{fig:compression}).

In between these two velocity limits, again as in the equal phase case, the collision does not yield a final black hole. Instead, it results in the two OS bouncing back, and then dispersal. While the OS experience compression during the bounce, the compression is not sufficient to push the OS into an unstable regime that led to collapse -- instead it led to a dispersion of the OS into scalar waves. While oscillatons have been shown to be stable under large spherically symmetric (and shell-like) perturbations \cite{Alcubierre:2003sx}, the perturbations that OS here experience post-bounce are both highly asymmetric and non-shell-like. Thus our results strongly suggests that there exist unstable \emph{non-radial} perturbation modes of OS even at low compactness, although a more detailed study is needed to confirm this conjecture.

\section{Discussion}
\label{sec:discussion}

The most striking result of our simulations is the existence of a ``stability band'' of velocities whereby collisions of OS do not form black holes.  We can gain a qualitative understanding as follows. The free fall time scale is given by $\tau_{\mathrm{ff}} \sim 1/\sqrt{G\rho}$, and using $\rho \sim M/R^3$ combined with \keq{eqn:compactness} gives
\begin{equation}
\tau_{\mathrm{ff}} \sim \frac{GM}{\mathcal{C}^{3/2}}~. \label{eqn:freefall}
\end{equation}
Meanwhile the interaction timescale can be estimated by the time the two OS overlap since the scalar field configuration of the OS drop off exponentially away from its characteristic size $R$.  If we assume that OS ``pass through'' (or bounce back after contact), then roughly the interaction timescale is
\begin{equation}
\tau_{\mathrm{int}} \sim \frac{2R}{\gamma v}  = \frac{2GM}{\gamma v\mathcal{C}}~. \label{eqn:interaction}
\end{equation}
This a conservative (i.e.  \emph{lower}) bound on $\tau_{\mathrm{int}}$ since interactions do slow down the collision -- as we saw especially in the anti-phase case the repulsion slows the collision down significantly, saturating only in the high $v$ limit. 

To prevent black hole formation, as we argued in Section \ref{sect:boostedOS} the interaction timescale has to be shorter than the free-fall timescale $\tau_{\mathrm{int}} > \tau_{\mathrm{ff}}$. At low $v$, $\gamma \sim 1$, we obtained the following bound
\begin{equation}
v > 2\mathcal{C}^{1/2}~. \label{eqn:vlower}
\end{equation}
Since $\tau_{\mathrm{int}}$ is an underestimate, we expect \keq{eqn:vlower} to be a lower bound on $v$. Combining this with the reduced hoop conjecture limit at high $\gamma$ \keq{eqn:hoop_2}, we obtain the following bound when BHs will not form
\begin{equation}
2\mathcal{C}^{1/2}< v < \sqrt{1-144\mathcal{C}^2}~. \label{eqn:bound}
\end{equation}
The two lines intersect at $\mathcal{C}\sim 0.07$ or $v\sim 0.5$, which is what we found numerically (see Fig. \ref{fig:money}). On the other hand, the lower bound does not track the numerical results accurately -- this is not surprising since such timescales arguments do not capture the full range of physics involved. 

An interesting question is whether this point is a ``critical point'', in the sense that the two different regimes $v>2{\cal C}^{1/2}$ and $v<\sqrt{1-144\mathcal{C}^2}$ constitute different phases and this point is where they meet as they transition into the final black hole phase. 

Since the two regimes exhibit different post collision behavior, it is interesting to ask whether their respective end states are the same  or are they different? In other words, is there a transition in the end states between the high $v$ BH formation and low $v$ BH formation in the black hole phase when $\mathcal{C}\gtrsim 0.07$?  The natural end state for these collisions are spherical, non-rotating black holes, hence the no-hair theorem implies that their end states are fully quantified by their final BH masses. To obtain these values require running the simulations to sufficiently long timescales to achieve these final states in addition to removing the unwanted reflection of scalar and tensor waves from the boundary of the simulation domain. We are currently exploring absorptive boundary conditions to overcome this problem. We will leave this, and the computation of gravitational waves signal from such collisions to a future publication. 

\acknowledgments
We would like to thank Mustafa Amin, Marcos Garcia, Helvi Witek and Lam Hui for very useful discussions and Ricardo Becerril for the use of his initial condition code for oscillatons. We would also like to thank the members of the \textsc{GRChombo} Collaboration (http://www.grchombo.org/) and the COSMOS team at DAMTP, Cambridge University for their ongoing technical support. EL is supported by STFC AGP grant ST/P000606/1, and JW is supported by a STFC PhD studentship. TH is supported by NSF Grant No. PHY-1912550, NSF Grant No. AST-1841358, NSF-XSEDE Grant No. PHY-090003, and NASA ATP Grant No. 17-ATP17-0225. This work has received funding from the European Union’s Horizon 2020 research and innovation programme under the Marie Skłodowska-Curie grant agreement No. 690904. The authors would like to acknowledge networking support by the GWverse COST Action CA16104,  ``Black holes, gravitational waves and fundamental physics.'' Numerical simulations were performed on the COSMOS supercomputer, the Cambridge CSD3 Peta4 and the Leicester DiAL (Data Intensive Cluster), all funded by DIRAC/BIS, on BSC Marenostrum IV via PRACE grant Tier-0 PPF-PWG, and on Leibnitz Supercomputing Center SuperMUC-NG under PRACE grant Tier-0 Proposal 2018194669. The simulation results were analyzed using the visualization toolkit YT \kcite{2011ApJS..192....9T} and Numpy \kcite{scipy}. Matplotlib \kcite{Hunter:2007} was used to generate the plots seen throughout the paper.

\bibliographystyle{h-physrev3.bst}

\bibliography{paper}
\newpage
\appendix

\section{Numerics}

\subsection{Gauge choice}

$\textsc{GRChombo}$ uses the BSSN formalism \kcite{Baumgarte:1998te,Nakamura,Shibata:1995we} of the Einstein equation in 3+1 dimensions using the ADM variables. The 4 dimensional spacetime metric is decomposed into a spatial metric on a 3 dimensional spatial hypersurface, $\gamma_{ij}$, and an extrinsic curvature $K_{ij}$, which are both evolved along a chosen local time coordinate $t$. 
The line element of the decomposition is
\begin{equation}
ds^2=-\alpha^2\,dt^2+\gamma_{ij}(dx^i + \beta^i\,dt)(dx^j + \beta^j\,dt)\,,
\end{equation}
where $\alpha$ and $\beta^i$ are the lapse and shift, the gauge parameters which must be specified.  These parameters are specified on the initial hypersurface (see below) and then allowed to evolve using gauge-driver equations in order to response dynamically to the physical system to ensure stable numerical evolution. In this work, we employ a slight modification of the usual puncture gauge \kcite{Campanelli:2005dd,Baker:2005vv}
\begin{equation}
\begin{split}
\partial_t \alpha = & -\mu \alpha K + \beta^i \partial_i \alpha\\
& - \left\{\begin{array}{l}
\epsilon_1(\alpha -\alpha_{\mathrm{analytic}}(t)) \phantom{e^{-\left ( \frac{t-t_{cut}}{t_{decay}} \right )^2}} \,t<t_{\mathrm{merger}} \\
\epsilon_2(\alpha -\alpha_{\mathrm{constant}}) e^{-\left ( \frac{t-t_{\mathrm{merger}}}{t_{\mathrm{decay}}} \right )^2} \, \,t\geq t_{\mathrm{merger}}
\end{array}\right. \, ,
\end{split}
\label{eqn:modified_lapse}
\end{equation}
for the lapse and
\begin{equation}
\partial_t \beta^i = \frac{3}{4} \Gamma^i - \eta \beta^i \, ,
\label{eqn:modified_shift}
\end{equation}
for the shift.

Note that for  \keq{eqn:modified_lapse}, we return to the usual puncture gauge driver when $\epsilon_1=\epsilon_2=0$, with the constants $\eta\sim 1/M_{ADM}$ and $\mu \sim 1$ usually chosen simulation-wise to improve stability. The additional terms with $\epsilon_1=\epsilon_2=1$ are added in to control the presence of $\mathcal{O}(1)$ gauge waves which propagate in the direction of the boost from each OS.  This effect caused unwanted adapative mesh refinement of the grid, increasing inaccuracy.

To eliminate these gauge waves, we drive $\alpha$ as close to $\alpha_{\mathrm{analytic}}$, which is the \emph{initial functional form} of $\alpha$, but evaluated at time $t$. This is done by adding the $\epsilon_1$ friction term. After merger at $t_{\mathrm{merger}}$, the system is quickly driven back into the standard puncture gauge form, and hence we switch from the $\epsilon_1$ friction term to the $\epsilon_2$ friction term which will exponentially drive this additional contributions $\alpha$ into zero. The rate of this decay can be set, and we found $t_{\mathrm{decay}}=7$ gives stable results.

Finally, we have chosen to use the first order ``integrated'' shift driver condition \cite{PhysRevLett.116.071102,Figueras:2017zwa} \keq{eqn:modified_shift} (as opposed to the usual puncture gauge driver condition which is 2nd order in time \cite{Campanelli:2005dd,Baker:2005vv}).  We find that this condition eliminates static imprints of initial values with $\tilde{\gamma}_{ij} \neq \delta_{ij} $ which remain even after the OS has moved from its initial position.

\subsection{Constructing Initial data}
\label{subsect:constructing_initial_data}
We construct our initial data by solving for a single oscillaton (OS) profile \kcite{PhysRevLett.66.1659,Alcubierre:2003sx,UrenaLopez:2002gx,UrenaLopez:2001tw}, boosting it and then superimposing them as per the methodology presented in \kcite{Helfer:2018vtq}. 

To obtain the radial OS profiles we use the ansatz for the spherically symmetric line element:
\begin{equation}
ds^2 = -\alpha^2 dt^2+a^2 dr^2 + r^2(d\theta^2+\sin^2(\theta)d\phi^2),
\end{equation}
from which we can define $A \equiv a^2$ and $C \equiv a^2/\alpha^2$. Solutions are then obtained by expanding the metric functions and the scalar field in their Fourier components:

\begin{equation}
\label{evolutionOsc}
\begin{split}
\phi(t,r) &=  \sum_{j=1}^{j_{\rm max}} \phi_{j}(r) \cos\left({j\omega t} \right) \\
A(t,r) &= \sum_{j=0}^{j_{\rm max}} A_{j}(r) \cos\left({j\omega t} \right), \\
C(t,r) &= \sum_{j=0 }^{j_{\rm max}} C_{j}(r) \cos\left({j\omega t} \right),
\end{split}
\end{equation}
where $\omega$ is a coherently oscillating base frequency and $j_{\rm max}$ is the maximum order in the Fourier expansion to which the solution is obtained. We substitute the Fourier expansion into the Einstein-Klein-Gordon system of equations with $V(\phi)=m^2\phi^2/2$. The Fourier coefficients, and $\omega$, can be found numerically, using a shooting technique. To generate these solutions,  we iteratively solve this hierarchical set of equations up to $j_{\mathrm{max}}=12$ following \cite{Alcubierre_2003}.

To apply a boost in the positive x-direction with some velocity, $v$, we define a Lorentz transformation of
\begin{equation}
\label{eqn:lorentz_boost}
\begin{split}
t &= \gamma (t' + v x'), \\
x  &= \gamma (x' + v t'), \\
y &= y', \\
z &= z',
\end{split}
\end{equation}
where $\gamma \equiv (1-v^2)^{-\frac{1}{2}}$ is the Lorentz factor. We denote $(t,x,y,z)$ as the ``lab" frame coordinates (the coordinates that relate to the Cartesian coordinates in \textsc{GRChombo}, and ultimately what we construct our initial conditions in), and  $(t',x',y',z')$ as the coordinates in the OS ``rest" frame.

We Lorentz boost both $A$, $C$, $\phi$, as well as the metric components. For an OS pair at rest the solutions for $A$, $C$, $\phi$ are taken on a hyperslice at $t_i =  \theta/\omega$, where we define $\theta$ as $=0$ for one OS, and $\theta=\Delta \theta$ for the second. For boosted OS we define the symmetry between them during superposition, and we choose that $A$, $C$, and $\phi$ are taken on a hyperslice  with $t=0$.

Given this single OS profile, we generate static OS-OS initial data by superposing two single OS solutions:
\begin{equation}
\begin{split}
\phi_{\rm tot} &= \phi|_{x-x_0}\pm \phi|_{x+x_0}\\
\pi_{\rm tot} &= \pi|_{x+x_0}\pm\pi|_{x-x_0} \\
\gamma_{ij,{\rm tot}} &= \gamma_{ij}|_{x+x_0}+ \gamma_{ij}|_{x-x_0} - h_{ij} \\
\end{split},
\end{equation}
where $\pm x_0$ are the locations of the centers of the two OS, and $h_{ij}$ is a constant metric. The $\pm$ relates the the overall symmetry of the OS pair, with $+$ relating to a symmetric arrangement and a $-$ relating to an antisymmetric configuration. For unboosted OS, as stated before, the symmetry is defined by which initial $t$ is chosen for each star, and a $+$ is used when superimposing. We then define $h_{ij} = \gamma_{ij}|_{2x_0}$ such that each OS is unchanged from its isolated case such that we obtained two \emph{unexcited} initial OS. If instead we had used $h_{ij}=\delta_{ij}$, the constructed OS possess significant radial modes which will lead to premature collapse of the OS into BH. See \kcite{Helfer:2018vtq} for a full discussion.

The linear superposition of the metric causes some violation of the Hamiltonian constraint. We quantify this violation using the relative constraint violation
\begin{equation}
\mathcal{H} \equiv \frac{H_{\mathrm{center}}}{16 \pi G \rho_{\mathrm{center}}} \,
\end{equation}
and find values of $\mathcal{O}(0.02) \%$ for an unboosted symmetric OS pair. The relative momentum constraint violation can be defined in a similar way. When an OS pair is boosted we use a relaxation routine in $\chi$ and $A_{ij}$ to reduce the relative violation in the Hamiltonian constraint and momentum constraint violation, as without this relaxation, the relative constraints would be $\mathcal{O}\approx 1\%$.

\subsection{Convergence and Stability}
\label{subsect:convergence}
We use the following to measure the volume averaged Hamiltonian constraint violation:
\begin{equation}\label{eq:L2H}
L^2(H) =\sqrt{\frac{1}{V} \int_V |\mathcal{H}^2| dV},
\end{equation}
where $V$ is the box volume with the interior of the apparent horizon excised. The volume averaged momentum constraint violation is calculated in a similar manner:
\begin{equation}\label{eq:L2H_2}
L^2(M) =\sqrt{\frac{1}{V} \int_V |\mathcal{M}^2| dV},
\end{equation}
We have good control over the constraint violation throughout our simulations, with a bouncing unboosted antisymmetric OS collision achieving a maximum value of $\mathcal{O}(10^{-6})$ at the beginning of the simulation and then decaying throughout the remainder of the simulation.

\begin{figure}
\centering
\includegraphics[width=1\columnwidth]{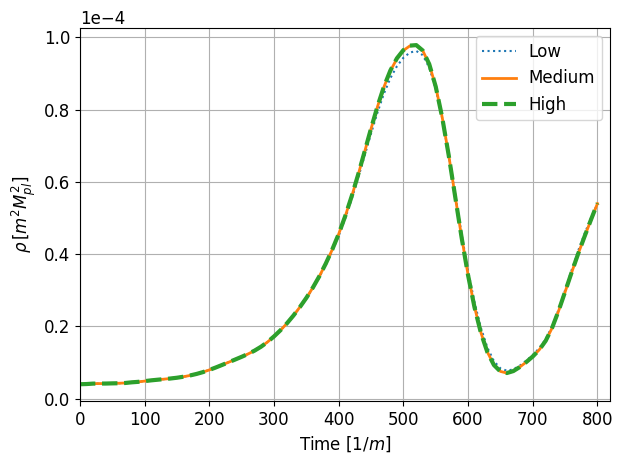}
\caption{The value of $\rho$ for a point along the collision axis of an antisymmetric OS pair that bounces. The test was done with fixed grid with three different resolutions of 1.0 $m^{-1}$, 0.5 $m^{-1}$ and 0.25 $m^{-1}$. We obtain an order of convergence between 3rd and 4th order for this test.}
\label{fig:convergence}
\end{figure}

We test the convergence of our simulations by measuring the value of $\rho$ along the collision axis of an unboosted antisymmetric OS pair with initial compactness of $\mathcal{C}=0.028$, that results in a bounce. The spatial coordinates for the value to be measured at was chosen such that the OS passes through it before and after it bounces. We used fixed grid for the convergence test with resolutions of 1.0 $m^{-1}$, 0.5 $m^{-1}$ and 0.25 $m^{-1}$. \kfig{fig:convergence} shows the value of $\rho$ for this test, and when calculated we obtain an order of convergence between 3rd and 4th order.

\end{document}